\documentclass[sigconf]{acmart}
\usepackage{amsmath}

\AtBeginDocument{%
  \providecommand\BibTeX{{%
    \normalfont B\kern-0.5em{\scshape i\kern-0.25em b}\kern-0.8em\TeX}}}

\setcopyright{acmcopyright}
\copyrightyear{2020}
\acmYear{2020}

\acmConference[SIGIR '20]{SIGIR '20: ACM Conference on Research and Development in Information Retrieval }{July 25-30, 2020}{Xi'an, China}
\acmBooktitle{}
\acmPrice{}
\acmISBN{}

\acmSubmissionID{1257}

\begin{document}

\title{Distilling Knowledge for Fast Retrieval-based Chat-bots}

\author{Amir Vakili Tahami}
\email{a_vakili@ut.ac.ir}
\affiliation{%
  \institution{University of Tehran}
}

\author{Kamyar Ghajar}
\email{k.ghajar@ut.ac.ir}
\affiliation{%
  \institution{University of Tehran}
}

\author{Azadeh Shakery}
\email{shakery@ut.ac.ir}
\affiliation{%
  \institution{University of Tehran}
}


\begin{abstract}
Response retrieval is a subset of neural ranking in which a model selects a suitable response from a set of candidates given a conversation history. Retrieval-based chat-bots are typically employed in information seeking conversational systems such as customer support agents.
In order to make pairwise comparisons between a conversation history and a candidate response, two approaches are common: cross-encoders performing full self-attention over the pair and bi-encoders encoding the pair separately. The former gives better prediction quality but is too slow for practical use. In this paper, we propose a new cross-encoder architecture and transfer knowledge from this model to a bi-encoder model using distillation. This effectively boosts bi-encoder performance at no cost during inference time. We perform a detailed analysis of this approach on three response retrieval datasets.
\end{abstract}




\maketitle

\section{Introduction}

Response retrieval is a subset of neural ranking in which a model selects a suitable response from a set of candidates given a conversation history. Retrieval-based chat-bots are typically employed in information seeking conversational systems such as customer support agents. They have been used in real-world products such as Microsoft XiaoIce \cite{shum2018eliza} and Alibaba Group’s AliMe Assist \cite{feng2017alime}. 

To find the best response to a particular conversation's chat history traditional text retrieval methods such as term frequency have proven to be insufficient \cite{lowe2015ubuntu}, therefore the majority of modern research focuses on neural ranking approaches \cite{lowe2015ubuntu,humeau2019real,henderson2019convert}. These methods rely on training artificial neural networks on large datasets for the task of selecting a suitable response among a set of candidates according to a conversation history.

By pre-training large scale language models on vast corpora and subsequently fine-tuning these models on downstream tasks, researchers have achieved state-of-the-art results in a wide variety of natural language tasks \cite{devlin2019bert}. This process has also been successfully applied to the task of response retrieval \cite{penha2019curriculum,humeau2019real,henderson2019convert}. Current state-of-the-art response retrieval focuses on using these pre-trained transformer language models such as BERT \cite{devlin2019bert}. When using a deep pre-trained transformer for the task of comparing two text inputs, two approaches are common: either encoding representations separately (bi-encoding) or encoding the concatenation of the two (cross-encoding). The BERT bi-encoder encodes two separate representations using pre-trained deep multi-layer transformers and compares them using a dot product operation. The BERT cross-encoder concatenates the conversation history and candidate response and encodes them into a single representation, which is fed into a fully connected network that gives a matching score. The latter method achieves better prediction quality but is far too slow for practical use \cite{humeau2019real}. 


While bi-encoding does give worse results, previous work has shown that one can significantly reduce its inference time by pre-encoding candidate responses offline so that during inference, only the conversation history needs to be encoded. This, in turn, means that at inference time, bi-encoders can potentially perform pairwise comparisons between a conversation history and millions of candidate responses. Such a feat is impossible to do with cross-encoders as they must recalculate encodings for each conversation history and candidate response pair. Naturally, this makes bi-encoders a desirable solution in conversational systems where real-time response selection is required \cite{humeau2019real}. Because of this improving the performance of bi-encoders is a popular avenue of research when it comes to response retrieval.




In this paper, we demonstrate one possible improvement to bi-encoders, which will boost their prediction quality without affecting their prediction speed.
We propose transferring knowledge from the better performing BERT cross-encoder to the much faster BERT bi-encoder. This method will raise BERT bi-encoder prediction quality without increasing inference time. We employ knowledge distillation, which is an approach where a model teaches another model to mimic it as a student \cite{hinton2015distilling}. Essentially, the student model learns to reproduce the outputs of the more complex teacher model. Unlike gold labels, the output of a neural network is not constrained to a binary variable and as such it can provide a much richer signal when training the student model. Knowledge distillation has been successfully applied in natural language understanding, machine translation, and language modeling tasks \cite{tang2019distilling,kim2016sequence,yu2018device}. 



We also introduce a new cross-encoder architecture we call the enhanced BERT cross-encoder. This architecture is specifically designed for the task of response retrieval and gives better results than the regular BERT cross-encoder. It also has the advantage of being faster to train. This model serves as our teacher, and we use the BERT bi-encoder \cite{humeau2019real} as our student model.
We evaluate our approach on three response retrieval data-sets. Our experiments show that our knowledge distillation approach  enhances the prediction quality of BERT the bi-encoder. This increase comes to a no-cost during inference time.

\begin{table}
\footnotesize
\caption{ Statistics for the datasets.}\label{tab:stats}
\centering
\begin{tabular}{lllllllllllll}
        \toprule
         & 
         \multicolumn{3}{c}{UDC} & \multicolumn{3}{c}{DSTC7} & \multicolumn{3}{c}{MANtIS}\\ 
         \cmidrule(lr){2-4}\cmidrule(lr){5-7}\cmidrule(lr){8-10}
         \textnumero \ of candidates &  \multicolumn{3}{c}{10} &  \multicolumn{3}{c}{100} & \multicolumn{3}{c}{11}   \\
         \midrule
            &  Trn & Vld  & Tst &  Trn & Vld  & Tst &  Trn & Vld  & Tst \\
        \midrule
        \textnumero \ of samples &    500k  &  50k  & 50k & 100k & 5k & 1k & 82k & 18k   & 18k   \\
        \bottomrule
    \end{tabular}
\end{table}

\section{Method}
First, we explain the task in further detail. Next, we describe the teacher and student models used for the knowledge distillation approach. Then we describe the knowledge distillation procedure.

\subsection{Task Definition}
The  task  of response retrieval can  be  formalized  as  follows:  Suppose  we  have  a  dataset $\mathcal{D}=\{c_i, r_i, y_i\}^N_{i=1}$ where $c_i=\{t_1,\cdots, t_m\}$ represents the conversation and $r_i=\{t_1,\cdots, t_n\}$ represents a candidate response and $y_i \in \{0,1\}$ is a label. $y_i= 1$ means that $r_i$ is a suitable choice for $c_i$. ${t_i}$ are tokens extracted from text . The goal of a model should be to learn a function $g(c, r)$ that predicts the matching degree between any new conversation history $c$ and a candidate response $r$. Once a given model ranks a set of candidates, its prediction quality is then measured using recall@1 (1 if the model's first choice is correct otherwise 0) and mean reciprocal rank (MRR).

\subsection{Model Architecture}
For the student network, we use the previously proposed BERT bi-encoder \cite{humeau2019real}. 
The conversation history and response candidate tokens are encoded separately using BERT. To aggregate the final layer encodings into a single vector, the first token's encoding, which corresponds to an individual [CLS] token, is selected. BERT requires all inputs to be prepended with this special token. The two aggregated vectors are compared using a dot-product operation.

Similarly, our teacher model uses a BERT transformer to encode the conversation history and candidate response. However, for comparing the last layer encodings we use a combination of scaled dot-product attention \cite{vaswani2017attention} and the \textit{SubMult} function \cite{wang2017compagg} for calculating the matching score. Below we give a brief explanation of these components before describing how they are used.

 In an attention mechanism, each entry of a key vector $k\in\mathbb{R}^{n_k\times d}$ is weighted by an importance score defined by its similarity to each entry of query $q\in\mathbb{R}^{n_q\times d}$. For each entry of $q$ the entries of $k$ are then linearly combined with the weights to form a new representation. Scaled dot-product attention is a particular version of attention defined as:

\begin{equation}
Att(q,k) = softmax\left(\frac{q \cdot k^T}{ \sqrt{d}}\right) \cdot k    
\end{equation}{}

The \textit{SubMult} function \cite{wang2017compagg} is a function designed for comparing two vectors $a\in\mathbb{R}^{d}$ and $b\in\mathbb{R}^{d}$ which has been used to great effect in various text matching tasks including response retrieval \cite{tao2019multi}. It is defined as follows:

\begin{equation}
SubMult(a,b) = a \oplus b \oplus (a - b) \oplus (a \odot b)    
\end{equation}{}

where $\oplus$ and $\odot$ are concatenation and hadamard product operators respectively.

Utilizing these components we build our enhanced cross-encoder architecture. First, like the bi-encoder, we encode the conversation history $c\in \mathbb{R}^{m\times d}$ and candidate response $r\in \mathbb{R}^{n\times d}$ as follows:

$$
\begin{array}{cc}
c' = T(c) \ , \  r' = T(r)
\end{array}{}
$$

where $T$ is the BERT transformer and $c'\in \mathbb{R}^{m \times d}, r'\in \mathbb{R}^{n \times d}$ are the encoded tokens. 

To compare the encoded conversation history $c'$ and encoded candidate response $r'$, first we perform a cross attention operation using the previously described components: 

\begin{equation} \label{eq:crossatt}
\begin{array}{cc}
     \hat{c} &= W_1 \cdot SubMult(c', Att(c',r'))  \\
     \hat{r} &= W_1 \cdot SubMult(r', Att(r',c'))  \\
\end{array}{}
\end{equation}

where $W_1\in\mathbb{R}^{4d\times d}$ is a a learned parameter.
We aggregate $\hat{c}\in \mathbb{R}^{m \times d}$ and $\hat{r} \in \mathbb{R}^{n \times d}$ by concatenating the first token (corresponding to [CLS]), the max pool and average pool over the tokens:

\begin{equation} \label{eq:agg}
\begin{array}{cc}
    \bar{c} &= \hat{c}_1 \oplus \max\limits_{1 \leq i \leq m}\hat{c}_{i} \oplus \underset{1 \leq i \leq m}{\text{mean}} \ {\hat{c}_{i}}  \\
    \bar{r} &= \hat{r}_1 \oplus \max\limits_{1 \leq i \leq n}\hat{r}_{i} \oplus \underset{1 \leq i \leq n}{\text{mean}} \ {\hat{r}_{i}}
\end{array}{}    
\end{equation}{}

We compare the aggregated $\bar{c},\bar{r}\in\mathbb{R}^d$ vectors using a final \textit{SubMult} function and a two layer fully connected network:
$$
g(c,r) =  W_2(ReLU(W_3 \cdot SubMult(\bar{c}, \bar{r})))
$$
where $W_2\in\mathbb{R}^{12d\times d}, W_3\in\mathbb{R}^{d\times 1}$ are learned parameters. Our enhanced BERT architecture essentially encodes the conversation history and candidate response tokens separately using BERT, then applies as single layer of cross-attention on those encodings.

We believe our enhanced cross-encoder architecture will perform better than regular cross-encoders for two reasons. Firstly, we do not concatenate conversation history and candidate responses. This means we can use the encoded candidate response tokens of other samples in a training batch as negative samples \cite{mazare2018millions}. Scaled dot-product attention is simple enough that recalculating it for other candidates in the batch does not add significant overhead, especially when compared to rerunning BERT for every possible conversation history and candidate response pair. Thus we can process more negative samples than would be feasible in a regular cross-encoder. Previous research has already shown that increasing the number of negative samples is effective for response retrieval \cite{humeau2019real}.  Secondly, the addition of the \textit{SubMult} function means we can achieve much more refined text matching between the conversation history and candidate response.

\subsection{Distillation Objective} \label{sec:DO}
Distillation achieves knowledge transfer at the output level. The student learns from both dataset gold labels and teacher predicted probabilities, which are also a useful source of information \cite{ba2014deep}. For example, in sentiment classification, certain sentences might have very strong or weak polarities and binary labels are not enough to convey this information.

Similar to previous work \cite{tang2019distilling}, we add a distillation objective to our loss function which penalizes the mean squared error loss between the student and teacher model outputs:

$$
\mathcal{L}_{\text{distill}} = ||z^{(T)}-z^{(S)}||^2
$$

where $z^{(T)},z^{(S)}$ are the teacher and student model outputs. At training time the distillation objective is used in conjunction with regular cross entropy loss as follows:

$$
\begin{array}{cc}
\mathcal{L}&= \alpha \cdot \mathcal{L}_{CE} + (1-\alpha)\cdot \mathcal{L}_{\text{distill}} \\
\end{array}{}
$$
where $\alpha$ is a hyper-parameter. This procedure is model agnostic and can transfer information between entirely different architectures.

\begin{table*}
\caption{Prediction quality metrics across all datasets. Metrics for models trained with knowledge distillation, which are significant relative to models trained without it, are marked in bold. We use paired two-tailed t-tests with a p-value<0.05 to perform significance tests. For easier reading metrics have been multiplied by 100. No data augmentation has been used and training samples are used as is. +KD indicates a model trained with knowledge distillation.}\label{tab:results}
\centering
\begin{tabular}{lllllllllllll}

        \toprule
         & 
         \multicolumn{2}{c}{UDC\textsubscript{20\%}} & \multicolumn{2}{c}{UDC} & \multicolumn{2}{c}{MANtIS} & \multicolumn{2}{c}{DSTC7}\\ 
         \cmidrule(lr){2-3}\cmidrule(lr){4-5}\cmidrule(lr){6-7}\cmidrule(lr){8-9}
                                            &  R@1 & MRR  & R@1  & MRR  & R@1  & MRR  & R@1  & MRR     \\
                                            \midrule
        BERT cross                          & 66.1 & 76.8 & 76.5 & 84.8 & 59.8 & 72.0 & 36.9 & 47.9 \\
        BERT cross enhanced                 & \bf76.2 & \bf84.5 & \bf79.5 & \bf86.9 & \bf66.7 & \bf77.3 & \bf53.3 & \bf63.3 \\
         - SubMult & 73.4& 82.6 & --- & --- & --- & --- & --- & --- \\
         - Attention & 67.2 & 78.6 & --- & --- & --- & --- & --- & --- \\
         \midrule
         BiLSTM bi-encoder                  & 59.2 & 72.4 & 69.4 & 80.2 & 35.6 & 55.1 & 34.3 & 46.1 \\
         BiLSTM bi-encoder + KD             & \bf63.0 & \bf75.2 & \bf70.4 & 80.8 & \bf45.5 & \bf61.4 & \bf39.4 & \bf50.1 \\
         \midrule
         BERT bi-encoder                    & 64.9 & 76.9 & 72.9 & 82.7 & 47.9 & 58.4 & 39.9 & 51.8 \\
         BERT bi-encoder + KD                 & \bf66.1 & \bf77.6 & \bf75.8 & \bf84.6 & \bf53.4 & \bf67.3 & \bf53.8 & \bf54.7 \\

    \end{tabular}
\end{table*}

\section{Experiments}
In this section we give a brief overview of experiments settings.
\subsection{Datasets}
We consider three information-seeking conversation datasets widely used in the training of neural ranking models for response retrieval. The Ubuntu Dialogue Corpus (UDC) \cite{lowe2015ubuntu} and DSTC7 sentence selection track dataset \cite{dstc19task1} are collected form a chatroom dedicated to the support of the Ubuntu operating system. We also include a version of UDC where the training set has been reduced to 20\% so as to study the effects of limited training data. MANtIS \cite{penha2019curriculum} was built from conversations of 14 different sites of the Stack Exchange Network. The statistics for these datasets are provided in Table \ref{tab:stats}. Data augmentation, where each conversation is split into multiple samples, is a popular method in dialog research for boosting the performance of response retrieval models. In this paper, we refrain from using this approach as our focus is not beating state-of-the-art results but empirically demonstrating the effectiveness of knowledge distillation even in limited-resource settings.

\subsection{Baselines}
We divide our experiments into three parts. 1. Comparing the regular BERT cross-encoder and our enhanced BERT cross-encoder. Here we aim to demonstrate the superiority of our proposed cross-encoder architecture 2. Comparing the BERT bi-encoder with and without distillation. Here we wish to demonstrate the effectiveness of the knowledge distillation approach. 3. Finally, we also train a BiLSTM bi-encoder with and without distillation in order to confirm the distillation process works with shallow student models. The BiLSTM bi-encoder uses the same tokens as BERT models, but their embeddings are not pre-trained and initialized randomly. We use the same aggregation strategy (eq. \ref{eq:agg}) to aggregate the BiLSTM hidden states. Our code will be released as open-source.

\subsection{Implementation Details}
Our models are implemented in the PyTorch framework \cite{paszke2017pytorch}. For our BERT component, we used Distilbert \cite{sanh2019distilbert} since it provides results somewhat close to the original implementation despite having only 6 layers of transformers instead of 12. We tune $\alpha$ from a set of $\{0.25, 0.5, 0.75\}$. We train models using Adam optimizer \cite{kingma2014adam}. We use a learning rate of $5\times10^{-5}$ for BERT models and $10^{-3}$ for the BiLSTM bi-encoder. For consistency, we set the batch size to 8 for all models. For each dataset, we set the maximum number of tokens in the conversation history and candidate responses so that no more than 20\% of inputs are truncated.

Unfortunately, due to limited computing resources, we are unable to beat state-of-the-art results reported by \cite{humeau2019real}. Our models are trained on a single GPU; thus, we had to make compromises on the number of input tokens, number of negative samples, and model depth.

\section{Results and Discussion}
In this section, we go over the results of our experiments. We analyze both prediction quality and efficiency.

\subsection{Prediction Quality}
 The first two rows of table \ref{tab:results} demonstrate the effectiveness our the enhanced BERT cross-encoder relative to the regular BERT cross-encoder. These results indicate that employing a task-specific single layer cross-attention mechanism on top of separately encoded inputs is highly effective for the task of response retrieval. Of particular note is the increased gap between the performance of the two methods when using smaller training sets (UDC\textsubscript{20\%}, MANtIS, DSTC7). This shows that the regular bert-cross model struggles when fine-tuned with smaller response-retrieval sets and data augmentation or a some other method must be used to achieve acceptable results. In contrast, our enhanced BERT cross-encoder's R@1 only dropped by 3.3 points when its training set was reduced to a fifth.
 
 To further demonstrate the effectiveness of our modifications to the BERT cross-encoder architecture, we perform an ablation study on the reduced UDC dataset. We replace the \textit{SubMult} function with a concatenation operation. We also try removing cross-attention (\ref{eq:crossatt}). In both cases, their removal significantly degrades model quality.

Across the datasets, bi-encoders show significant gains when trained with knowledge distillation. The increase in performance is relatively substantial. Such gains usually require an increase in model complexity, however with knowledge distillation, we are effectively gaining a free boost in performance as there is no extra cost at inference time. The best results were obtained with an $\alpha$ of 0.5. This indicates that in response retrieval, unlike other tasks such as sentiment classification and natural language inference \cite{tang2019distilling}, the gold labels cannot be replaced entirely with teacher outputs.

\subsection{Prediction Efficiency}
We demonstrate the trade-off in speed and performance between the BERT bi-encoder and our enhanced BERT cross-encoder. We measure the time it takes to process test samples in the DSTC7 dataset and show the average time for each example in table \ref{tab:speed}. Time taken by the cross-encoder to process a set of candidate responses grows exponentially large as the set increases in size. In the case of BERT bi-encoders, since candidate vectors can be computed offline, increasing candidates has a negligible impact on inference time.

\begin{table}
\caption{ Average milliseconds to process a single test sample.}\label{tab:speed}
\centering
\begin{tabular}{lll}
        \toprule
        \textnumero \ of candidates & 10 & 100 \\
        \midrule
BERT bi-encoder & 5.6 & 6.2\\
BERT cross-encoder enhanced & 81.1 & 981.2\\
        \bottomrule
    \end{tabular}
\end{table}

\section{Conclusion and Future Work}
In this paper, we introduced an enhanced BERT cross-encoder architecture modified for the task of response retrieval. Alongside that, we utilized knowledge distillation to compress the complex BERT cross-encoder network as a teacher model into the student BERT bi-encoder model. This increases the BERT bi-encoders prediction quality without affecting its inference speed. We evaluate our approach on three domain-popular datasets. The proposed methods were shown to achieve statistically significant gains. 

One possible avenue for research is the exploration of other knowledge transfer methods. Substituting the relatively simple BERT bi-encoder architecture with a more complex architecture \cite{henderson2019convert} or developing further improvements to the BERT cross-encoder are also viable alternatives.


\bibliographystyle{ACM-Reference-Format}
\bibliography{sample-authordraft}


\begin{thebibliography}{00}


\ifx \showCODEN    \undefined \def \showCODEN     #1{\unskip}     \fi
\ifx \showDOI      \undefined \def \showDOI       #1{#1}\fi
\ifx \showISBNx    \undefined \def \showISBNx     #1{\unskip}     \fi
\ifx \showISBNxiii \undefined \def \showISBNxiii  #1{\unskip}     \fi
\ifx \showISSN     \undefined \def \showISSN      #1{\unskip}     \fi
\ifx \showLCCN     \undefined \def \showLCCN      #1{\unskip}     \fi
\ifx \shownote     \undefined \def \shownote      #1{#1}          \fi
\ifx \showarticletitle \undefined \def \showarticletitle #1{#1}   \fi
\ifx \showURL      \undefined \def \showURL       {\relax}        \fi
\providecommand\bibfield[2]{#2}
\providecommand\bibinfo[2]{#2}
\providecommand\natexlab[1]{#1}
\providecommand\showeprint[2][]{arXiv:#2}

\bibitem[\protect\citeauthoryear{Ba and Caruana}{Ba and Caruana}{2014}]%
        {ba2014deep}
\bibfield{author}{\bibinfo{person}{Jimmy Ba} {and} \bibinfo{person}{Rich
  Caruana}.} \bibinfo{year}{2014}\natexlab{}.
\newblock \showarticletitle{Do deep nets really need to be deep?}. In
  \bibinfo{booktitle}{{\em Advances in neural information processing systems}}.
\newblock


\bibitem[\protect\citeauthoryear{Chulaka~Gunasekara and
  Lasecki}{Chulaka~Gunasekara and Lasecki}{2019}]%
        {dstc19task1}
\bibfield{author}{\bibinfo{person}{Lazaros~Polymenakos Chulaka~Gunasekara,
  Jonathan K.~Kummerfeld} {and} \bibinfo{person}{Walter~S. Lasecki}.}
  \bibinfo{year}{2019}\natexlab{}.
\newblock \showarticletitle{DSTC7 Task 1: Noetic End-to-End Response
  Selection}. In \bibinfo{booktitle}{{\em 7th Edition of the Dialog System
  Technology Challenges at AAAI 2019}}.
\newblock


\bibitem[\protect\citeauthoryear{Devlin, Chang, Lee, and Toutanova}{Devlin
  et~al\mbox{.}}{2019}]%
        {devlin2019bert}
\bibfield{author}{\bibinfo{person}{Jacob Devlin}, \bibinfo{person}{Ming-Wei
  Chang}, \bibinfo{person}{Kenton Lee}, {and} \bibinfo{person}{Kristina
  Toutanova}.} \bibinfo{year}{2019}\natexlab{}.
\newblock \showarticletitle{BERT: Pre-training of Deep Bidirectional
  Transformers for Language Understanding}. In \bibinfo{booktitle}{{\em
  Proceedings of the 2019 Conference of the North American Chapter of the
  Association for Computational Linguistics: Human Language Technologies}}.
\newblock


\bibitem[\protect\citeauthoryear{Henderson, Casanueva, Mrk{\v{s}}i{\'c}, Su,
  Vuli{\'c}, et~al\mbox{.}}{Henderson et~al\mbox{.}}{2019}]%
        {henderson2019convert}
\bibfield{author}{\bibinfo{person}{Matthew Henderson},
  \bibinfo{person}{I{\~n}igo Casanueva}, \bibinfo{person}{Nikola
  Mrk{\v{s}}i{\'c}}, \bibinfo{person}{Pei-Hao Su}, \bibinfo{person}{Ivan
  Vuli{\'c}}, {et~al\mbox{.}}} \bibinfo{year}{2019}\natexlab{}.
\newblock \showarticletitle{ConveRT: Efficient and Accurate Conversational
  Representations from Transformers}.
\newblock \bibinfo{journal}{{\em arXiv preprint arXiv:1911.03688\/}}
  (\bibinfo{year}{2019}).
\newblock


\bibitem[\protect\citeauthoryear{Hinton, Vinyals, and Dean}{Hinton
  et~al\mbox{.}}{2015}]%
        {hinton2015distilling}
\bibfield{author}{\bibinfo{person}{Geoffrey Hinton}, \bibinfo{person}{Oriol
  Vinyals}, {and} \bibinfo{person}{Jeff Dean}.}
  \bibinfo{year}{2015}\natexlab{}.
\newblock \showarticletitle{Distilling the knowledge in a neural network}.
\newblock \bibinfo{journal}{{\em arXiv preprint arXiv:1503.02531\/}}
  (\bibinfo{year}{2015}).
\newblock


\bibitem[\protect\citeauthoryear{Humeau, Shuster, Lachaux, and Weston}{Humeau
  et~al\mbox{.}}{2020}]%
        {humeau2019real}
\bibfield{author}{\bibinfo{person}{Samuel Humeau}, \bibinfo{person}{Kurt
  Shuster}, \bibinfo{person}{Marie-Anne Lachaux}, {and} \bibinfo{person}{Jason
  Weston}.} \bibinfo{year}{2020}\natexlab{}.
\newblock \showarticletitle{Poly-encoders: architectures and pre-training
  strategies for fast and accurate multi-sentence scoring}. In
  \bibinfo{booktitle}{{\em 8th International Conference on Learning
  Representations, ICLR 2020}}.
\newblock


\bibitem[\protect\citeauthoryear{Kim and Rush}{Kim and Rush}{2016}]%
        {kim2016sequence}
\bibfield{author}{\bibinfo{person}{Yoon Kim} {and} \bibinfo{person}{Alexander~M
  Rush}.} \bibinfo{year}{2016}\natexlab{}.
\newblock \showarticletitle{Sequence-Level Knowledge Distillation}. In
  \bibinfo{booktitle}{{\em Proceedings of the 2016 Conference on Empirical
  Methods in Natural Language Processing}}.
\newblock


\bibitem[\protect\citeauthoryear{Kingma and Ba}{Kingma and Ba}{2015}]%
        {kingma2014adam}
\bibfield{author}{\bibinfo{person}{Diederik~P. Kingma} {and}
  \bibinfo{person}{Jimmy Ba}.} \bibinfo{year}{2015}\natexlab{}.
\newblock \showarticletitle{Adam: {A} Method for Stochastic Optimization}. In
  \bibinfo{booktitle}{{\em 3rd International Conference on Learning
  Representations, {ICLR} 2015, San Diego, CA, USA, May 7-9, 2015, Conference
  Track Proceedings}}.
\newblock


\bibitem[\protect\citeauthoryear{Li, Qiu, Chen, Wang, Gao, Huang, Ren, Zhao,
  Zhao, Wang, Jin, and Chu}{Li et~al\mbox{.}}{2017}]%
        {feng2017alime}
\bibfield{author}{\bibinfo{person}{Feng{-}Lin Li}, \bibinfo{person}{Minghui
  Qiu}, \bibinfo{person}{Haiqing Chen}, \bibinfo{person}{Xiongwei Wang},
  \bibinfo{person}{Xing Gao}, \bibinfo{person}{Jun Huang},
  \bibinfo{person}{Juwei Ren}, \bibinfo{person}{Zhongzhou Zhao},
  \bibinfo{person}{Weipeng Zhao}, \bibinfo{person}{Lei Wang},
  \bibinfo{person}{Guwei Jin}, {and} \bibinfo{person}{Wei Chu}.}
  \bibinfo{year}{2017}\natexlab{}.
\newblock \showarticletitle{\emph{AliMe Assist }: An Intelligent Assistant for
  Creating an Innovative E-commerce Experience}. In \bibinfo{booktitle}{{\em
  Proceedings of the 2017 {ACM} on Conference on Information and Knowledge
  Management, {CIKM} 2017}}.
\newblock


\bibitem[\protect\citeauthoryear{Lowe, Pow, Serban, and Pineau}{Lowe
  et~al\mbox{.}}{2015}]%
        {lowe2015ubuntu}
\bibfield{author}{\bibinfo{person}{Ryan Lowe}, \bibinfo{person}{Nissan Pow},
  \bibinfo{person}{Iulian Serban}, {and} \bibinfo{person}{Joelle Pineau}.}
  \bibinfo{year}{2015}\natexlab{}.
\newblock \showarticletitle{The Ubuntu Dialogue Corpus: A Large Dataset for
  Research in Unstructured Multi-Turn Dialogue Systems}. In
  \bibinfo{booktitle}{{\em Proceedings of the 16th Annual Meeting of the
  Special Interest Group on Discourse and Dialogue}}.
\newblock


\bibitem[\protect\citeauthoryear{Mazar{\'{e}}, Humeau, Raison, and
  Bordes}{Mazar{\'{e}} et~al\mbox{.}}{2018}]%
        {mazare2018millions}
\bibfield{author}{\bibinfo{person}{Pierre{-}Emmanuel Mazar{\'{e}}},
  \bibinfo{person}{Samuel Humeau}, \bibinfo{person}{Martin Raison}, {and}
  \bibinfo{person}{Antoine Bordes}.} \bibinfo{year}{2018}\natexlab{}.
\newblock \showarticletitle{Training Millions of Personalized Dialogue Agents}.
  In \bibinfo{booktitle}{{\em Proceedings of the 2018 Conference on Empirical
  Methods in Natural Language Processing, Brussels, Belgium, October 31 -
  November 4, 2018}}.
\newblock


\bibitem[\protect\citeauthoryear{Paszke, Gross, Chintala, Chanan, Yang, DeVito,
  Lin, Desmaison, Antiga, and Lerer}{Paszke et~al\mbox{.}}{2017}]%
        {paszke2017pytorch}
\bibfield{author}{\bibinfo{person}{Adam Paszke}, \bibinfo{person}{Sam Gross},
  \bibinfo{person}{Soumith Chintala}, \bibinfo{person}{Gregory Chanan},
  \bibinfo{person}{Edward Yang}, \bibinfo{person}{Zachary DeVito},
  \bibinfo{person}{Zeming Lin}, \bibinfo{person}{Alban Desmaison},
  \bibinfo{person}{Luca Antiga}, {and} \bibinfo{person}{Adam Lerer}.}
  \bibinfo{year}{2017}\natexlab{}.
\newblock \showarticletitle{Automatic Differentiation in {PyTorch}}. In
  \bibinfo{booktitle}{{\em NIPS Autodiff Workshop}}.
\newblock


\bibitem[\protect\citeauthoryear{Penha and Hauff}{Penha and Hauff}{2020}]%
        {penha2019curriculum}
\bibfield{author}{\bibinfo{person}{Gustavo Penha} {and}
  \bibinfo{person}{Claudia Hauff}.} \bibinfo{year}{2020}\natexlab{}.
\newblock \showarticletitle{Curriculum Learning Strategies for IR: An Empirical
  Study on Conversation Response Ranking}. In \bibinfo{booktitle}{{\em European
  Conference on Information Retrieval}}. Springer.
\newblock


\bibitem[\protect\citeauthoryear{Sanh, Debut, Chaumond, and Wolf}{Sanh
  et~al\mbox{.}}{2019}]%
        {sanh2019distilbert}
\bibfield{author}{\bibinfo{person}{Victor Sanh}, \bibinfo{person}{Lysandre
  Debut}, \bibinfo{person}{Julien Chaumond}, {and} \bibinfo{person}{Thomas
  Wolf}.} \bibinfo{year}{2019}\natexlab{}.
\newblock \showarticletitle{DistilBERT, a distilled version of BERT: smaller,
  faster, cheaper and lighter}.
\newblock \bibinfo{journal}{{\em arXiv preprint arXiv:1910.01108\/}}
  (\bibinfo{year}{2019}).
\newblock


\bibitem[\protect\citeauthoryear{Shum, He, and Li}{Shum et~al\mbox{.}}{2018}]%
        {shum2018eliza}
\bibfield{author}{\bibinfo{person}{Heung-Yeung Shum},
  \bibinfo{person}{Xiao-dong He}, {and} \bibinfo{person}{Di Li}.}
  \bibinfo{year}{2018}\natexlab{}.
\newblock \showarticletitle{From Eliza to XiaoIce: challenges and opportunities
  with social chatbots}.
\newblock \bibinfo{journal}{{\em Frontiers of Information Technology \&
  Electronic Engineering\/}} (\bibinfo{year}{2018}).
\newblock


\bibitem[\protect\citeauthoryear{Tang, Lu, Liu, Mou, Vechtomova, and Lin}{Tang
  et~al\mbox{.}}{2019}]%
        {tang2019distilling}
\bibfield{author}{\bibinfo{person}{Raphael Tang}, \bibinfo{person}{Yao Lu},
  \bibinfo{person}{Linqing Liu}, \bibinfo{person}{Lili Mou},
  \bibinfo{person}{Olga Vechtomova}, {and} \bibinfo{person}{Jimmy Lin}.}
  \bibinfo{year}{2019}\natexlab{}.
\newblock \showarticletitle{Distilling task-specific knowledge from BERT into
  simple neural networks}.
\newblock \bibinfo{journal}{{\em arXiv preprint arXiv:1903.12136\/}}
  (\bibinfo{year}{2019}).
\newblock


\bibitem[\protect\citeauthoryear{Tao, Wu, Xu, Hu, Zhao, and Yan}{Tao
  et~al\mbox{.}}{2019}]%
        {tao2019multi}
\bibfield{author}{\bibinfo{person}{Chongyang Tao}, \bibinfo{person}{Wei Wu},
  \bibinfo{person}{Can Xu}, \bibinfo{person}{Wenpeng Hu},
  \bibinfo{person}{Dongyan Zhao}, {and} \bibinfo{person}{Rui Yan}.}
  \bibinfo{year}{2019}\natexlab{}.
\newblock \showarticletitle{Multi-Representation Fusion Network for Multi-Turn
  Response Selection in Retrieval-Based Chatbots}. In \bibinfo{booktitle}{{\em
  Proceedings of the Twelfth ACM International Conference on Web Search and
  Data Mining}}.
\newblock


\bibitem[\protect\citeauthoryear{Vaswani, Shazeer, Parmar, Uszkoreit, Jones,
  Gomez, Kaiser, and Polosukhin}{Vaswani et~al\mbox{.}}{2017}]%
        {vaswani2017attention}
\bibfield{author}{\bibinfo{person}{Ashish Vaswani}, \bibinfo{person}{Noam
  Shazeer}, \bibinfo{person}{Niki Parmar}, \bibinfo{person}{Jakob Uszkoreit},
  \bibinfo{person}{Llion Jones}, \bibinfo{person}{Aidan~N Gomez},
  \bibinfo{person}{{\L}ukasz Kaiser}, {and} \bibinfo{person}{Illia
  Polosukhin}.} \bibinfo{year}{2017}\natexlab{}.
\newblock \showarticletitle{Attention is all you need}. In
  \bibinfo{booktitle}{{\em Advances in neural information processing systems}}.
\newblock


\bibitem[\protect\citeauthoryear{Wang and Jiang}{Wang and Jiang}{2017}]%
        {wang2017compagg}
\bibfield{author}{\bibinfo{person}{Shuohang Wang} {and} \bibinfo{person}{Jing
  Jiang}.} \bibinfo{year}{2017}\natexlab{}.
\newblock \showarticletitle{A Compare-Aggregate Model for Matching Text
  Sequences}. In \bibinfo{booktitle}{{\em 5th International Conference on
  Learning Representations, {ICLR} 2017, Toulon, France, April 24-26, 2017,
  Conference Track Proceedings}}.
\newblock


\bibitem[\protect\citeauthoryear{Yu, Kulkarni, Lee, and Kim}{Yu
  et~al\mbox{.}}{2018}]%
        {yu2018device}
\bibfield{author}{\bibinfo{person}{Seunghak Yu}, \bibinfo{person}{Nilesh
  Kulkarni}, \bibinfo{person}{Haejun Lee}, {and} \bibinfo{person}{Jihie Kim}.}
  \bibinfo{year}{2018}\natexlab{}.
\newblock \showarticletitle{On-device neural language model based word
  prediction}. In \bibinfo{booktitle}{{\em Proceedings of the 27th
  International Conference on Computational Linguistics: System
  Demonstrations}}.
\newblock


\end{thebibliography}

\end{document}